\documentclass[11pt,a4paper]{scrartcl}
\usepackage[round,authoryear]{natbib}

\usepackage[top=2.5cm, bottom=2.5cm, left=2.5cm, right=2.5cm]{geometry}

\usepackage{booktabs}
\usepackage{color}
\usepackage{latexsym}              
\usepackage{amsmath}               
\usepackage{amssymb}               
\usepackage{amsfonts}              
\usepackage{amsthm}                
\usepackage{multirow}
\usepackage{tikz}                  
\usetikzlibrary{arrows,positioning,shapes}
\usepackage{tcolorbox}

\usepackage[english]{babel}

\RequirePackage[%
  pdfstartview=FitH,%
  breaklinks=true,%
  bookmarks=true,%
  colorlinks=true,%
  linkcolor= blue,
  anchorcolor=blue,%
  citecolor=blue,
  filecolor=blue,%
  menucolor=blue,%
  urlcolor=blue%
  ]{hyperref}
  
  \AtBeginDocument{%
  \hypersetup{%
    pdfauthor={Florent Dewez, Benjamin Guedj, Vincent Vandewalle},%
      colorlinks = true,%
    urlcolor = blue,%
    linkcolor = blue,%
    citecolor = orange,%
    pdftitle={From industry-wide parameters to aircraft-centric on-flight inference: improving aeronautics performance prediction with machine learning - compiled: \today}%
  }
}

\graphicspath{{figures/}}

\usepackage[mathcal]{eucal}

\renewenvironment{proof}[1][\proofname]{{\bfseries #1.}}{\qed \\ }

\makeatother

\DeclareMathOperator*{\argmin}{arg\,min}

\newcommand{\R}{\mathbb{R}}
\newcommand{\E}{\mathbb{E}}
\def\FF{\mathrm{FF}}
\def\SAT{\mathrm{SAT}}
\def\SR{\mathrm{SR}}

\newtheorem{lem}{Lemma}
\newtheorem{remark}{Remark}

\begin{document}
\title{From industry-wide parameters to aircraft-centric on-flight inference: improving aeronautics performance prediction with machine learning}

\author{\textbf{Florent Dewez}\footnote{Modal project-team, Lille - Nord Europe research centre, Inria, France} \\ [2ex]
\textbf{Benjamin Guedj}\footnote{Modal project-team, Lille - Nord Europe research centre, Inria, France; Centre for Artificial Intelligence, Department of Computer Science, University College London, United Kingdom} \\ [2ex]
\textbf{Vincent Vandewalle}\footnote{Modal project-team, Lille - Nord Europe research centre, Inria; Université de Lille, France} \\ [2ex]
}
\date{}

\maketitle

\begin{abstract}
Aircraft performance models play a key role in airline operations, especially in planning a fuel-efficient flight. In practice, manufacturers provide guidelines which are slightly modified throughout the aircraft life cycle via the tuning of a single factor, enabling better fuel predictions. However this has limitations, in particular they do not reflect the evolution of each feature impacting the aircraft performance. Our goal here is to overcome this limitation.
The key contribution of the present article is to foster the use of machine learning to leverage the massive amounts of data continuously recorded during flights performed by an aircraft and provide models reflecting its actual and individual performance.
We illustrate our approach by focusing on the estimation of the drag and lift coefficients from recorded flight data. As these coefficients are not directly recorded, we resort to aerodynamics approximations. As a safety check, we provide bounds to assess the accuracy of both the aerodynamics approximation and the statistical performance of our approach.
We provide numerical results on a collection of machine learning algorithms. We report excellent accuracy on real-life data and exhibit empirical evidence to support our modelling, in coherence with aerodynamics principles.
\end{abstract}

\section{Introduction}

Efficient aircraft operations require the knowledge of current performance of the aircraft. The performance depends on many aerodynamic and engines features which evolve throughout the life cycle of the aircraft. For instance the degradation of the engines may modify the aircraft specific range up to 1.3\% per year if no engine replacement is carried out \citep{airbus2002}. Further aircraft physical features may change overtime due to accumulations of impurities on the surface (leading to an increase of the drag), rough or deformed surfaces, damaged seals, \dots \citep[see for instance][]{airbus2001}. In particular performance may differ from an aircraft to another one of the same type due to different cycles, flight hours or maintenance.

Given this deterioration, monitoring systems have been developed by the manufacturers \citep{krajcek2015}. Those systems measure the difference between expected behaviour, obtained through heavy numerical simulations and wind tunnel tests, and data recorded during very specific in-flight conditions. Given this difference, most manufacturers propose to modify a performance index to adjust the theoretical fuel consumption to the current one. This leads to more accurate fuel predictions when planning the flight operations. However according to \cite{krajcek2015}, such monitoring systems have limitations, \emph{e.g.} the incapacity of dissociating the influence of aerodynamic or engines features on the evolution of performance, since it is only described by the performance index.

In view of this, a flexible methodology designed to build models for features which reflect the current performance of an aircraft would be relevant to improve flight planning operations in the end. Such a methodology is proposed in the present paper.

The underlying idea of our approach is that the real performance of an aircraft should be reflected by its data recorded in recent flights. Here we consider data from the Quick Access Recorder which contain features of different types such as the altitude, the true airspeed or the engine power, sampled every second. To exploit this data, we propose to model statistically features of interest and to fit these models on the recorded data with off-the-shelf machine learning algorithms. The resulting estimators are then expected to take into account the actual performance of the aircraft, thus leading to a far more precise description of its performance.

To illustrate our method, we propose to model the drag and lift coefficients. It is common in the aeronautic literature to model the drag and lift forces through these coefficients, which quantify drag and lift independently from the wing size, airspeed and air density \cite[Chapter 2]{MC1995}. In particular, the coefficients are used to somewhat capture very complex phenomena such as friction and permit to deduce the lift-to-drag ratio \cite[Chapter 7]{L1985}, which plays a key role in assessing the performance of the aircraft.

The choice of these two aerodynamic coefficients is also motivated by the fact that neither of those coefficients is recorded by the aircraft (as a matter of fact, the drag and lift forces are also not recorded), highlighting the flexibility of this statistical approach.
Indeed this issue is bypassed by leveraging physical relationships to obtain approximated but explicit and deterministic formulas for the drag and lift coefficients. The statistical models are then fitted to approximated train data and their learning errors are computed on test sets. 
It is noteworthy that the models have to depend on features set by the user since they are aimed at being exploited by aeronautic softwares, such as those embedded in the Flight Management System of the aircraft. This explains in particular why physical models which may depend on other features, may not be used in such a setting.

This approach induces an additional error which we refer to as a physical approximation error, coming from the approximated data. To assess the prediction accuracy of the fitted models, this approximation error has to be taken into consideration. In a general setting within which we introduce our approach, we propose bounds for the mean absolute error and relative error between the true value of the output and the predicted value from the model. These bounds depend explicitly on the physical approximation and the learning errors and are then applied in the present aeronautic setting. Note that in a slightly different setting, the problem can be interpreted as with errors on the response variable \citep{Buonaccorsi1996}, which is a particular case of the general framework of errors-in-variables models \citep{SCH2016,fuller2009}. For such problems a statistical model is assumed on the distribution of the observed surrogate response variable given the unobserved response variable, for instance the additive measurement error model \citep{carroll1988transformation}.

A similar approach to model unobserved aerodynamic variables has been proposed in \cite{SHE2018}. In this paper, the authors aim at estimating the drag polar (\emph{i.e.} a specific quadratic model for the drag coefficient depending on the lift one) by using a stochastic total energy model. Their approach is based on a MCMC sampler to estimate posterior probability distributions of their parameters of interest. Similarly to our methodology, they exploit physical formulas to obtain approximate values for unobserved variables. Nevertheless neither the associated error nor its impact on the prediction accuracy are taken into account in their analysis.

At this stage, let us stress that our aim  is to propose a flexible methodology which is sufficiently generic to be used in any data-intensive engineering discipline. We show that this methodology can be straightforwardly applied to an aeronautic setting and that it leads to accurate predictions for two aerodynamic features, verifying in particular expected tendencies. This is illustrated with extensive numerical tests on recorded flight data. However let us note that comparing the performance resulting from our estimated models to those provided by the manufacturers is unfortunately out of reach: in general the manufacturers performance models are not publicly available (for commercial reasons) and therefore cannot be used for academic research.

The paper is organised as follows: in Section \ref{sec:abs_setting}, we first propose an abstract formulation of the problem of modelling a variable for which only approximated data are available. Lemmas \ref{lem:lem1} and \ref{lem:lem2} provide the above mentioned bounds for the prediction error. Section \ref{sec:appl} aims at specifying the aeronautic setting of interest. Let us mention that we restrict our study to the cruise phase for which physical approximation errors values are available, however our method is actually not limited to this particular phase. Numerical results based on real flight data are presented and discussed in Section \ref{sec:num_exp}, and the paper closes on avenues for future work in Section \ref{sec:conclusion}.

\section{Statistical modelling} \label{sec:abs_setting}

In this section, we consider a general setting where one aims at explaining a real-valued random variable $Y^\star \in \mathcal{Y} \subseteq \R$ through a function $f^\star$ depending on the vector $X \in \mathcal{X} \subseteq \R^{d_{\mathcal{X}}}$. We formulate this as the following regression problem:
\begin{equation} \label{eq:regr1}
    Y^\star = f^\star(X) + \varepsilon \; ,
\end{equation}
where $\varepsilon$ denotes a noise variable standing for unexplained determinants of $Y^\star$. Nevertheless, in our setting, $Y^\star$ is a latent variable: no direct observation for $Y^\star$ is available, turning the direct estimation of $f^\star$ impossible. Our idea here is to propose an estimator for a surrogate of $Y^\star$ for which data can be obtained. To do so, suppose that there exists a relationship between $Y^\star$ and observed variables contained in a vector $Z \in \mathcal{Z} \subseteq \R^{d_{\mathcal{Z}}}$, which can be observed together with $X$. More precisely, we suppose that there exists a known and explicit function $\varphi: \mathcal{Z} \longrightarrow \R$ such that
\begin{equation*}
    \E_{Y^\star, Z} \Big[ \big| Y^\star - \varphi \big(Z \big) \big| \Big] \leqslant r \; ,
\end{equation*}
where $r > 0$ is known, and we let $\E_A$ denote the expectation with respect to a random variable $A$. Thus the variable
\begin{equation*}
    Y := \varphi \left(Z \right)
\end{equation*}
can be considered as an approximation of $Y^\star$, coming from a physical formula for instance. The error $\eta : \mathcal{Y^\star} \times \mathcal{Z} \longrightarrow \R$ of this approximation is defined as follows,
\begin{equation*}
    \forall \, (y^\star, z) \in \mathcal{Y} \times \mathcal{Z}, \qquad\qquad \eta(y^\star, z) := y^\star - \varphi(z) \; ,
\end{equation*}
and will be named the \emph{physical approximation error}.
We consider then the following regression problem,
\begin{equation} \label{eq:final_regr}
   Y = f(X) + \epsilon \; .
\end{equation}
Note that $\varphi(Z)$ is actually a model for the latent variable $Y^\star$ but our aim is to model this variable via the input $X$ and not $Z$, motivating the problem \eqref{eq:final_regr}. For instance, such a requirement arises in industrial contexts where inputs of some specialist softwares may differ from the variables involved in physical equations.
If we assume that we have access to $n$ random observations $(x_i, z_i)$ (realisations from $X$ and $Z$), we can derive observations for $Y$ as follows,
\begin{equation*}
    \forall \,  i= 1, \dots, n, \qquad\qquad y_i := \varphi(z_i) \; ,
\end{equation*}
leading to a training set $\mathcal{D} := \left(x_i, y_i \right)_{i=1}^{n}$. Contrary to the original problem \eqref{eq:regr1} for which no training set is available, an estimator $\hat{f}$ for the model $f$ can be derived by solving the following minimisation problem:
\begin{equation*}
    \hat{f} \in \argmin_{g \in \mathcal{H}} \sum_{i=1}^n \ell \left( y_i, g(x_i) \right) \; ,
\end{equation*}
where the hypothesis class $\mathcal{H}$ and the loss function $\ell: \R^2 \longrightarrow \R$ are generic at this stage. For instance, one may consider the class of polynomials and the squared error loss.

Let us now upper bound the mean of the absolute value of the \emph{total error}, defined by $Y^\star - \hat{f}(X)$. In other words, the total error is the error between the unobserved variable $Y^\star$ and the predicted value $\hat{f}(X)$ given the training set $\mathcal{D}$. Note that the total error can be decomposed as follows:
\begin{equation} \label{eq:decomp_error}
    Y^\star - \hat{f}(X) = \eta(Y^\star, Z) + \Big(Y - \hat{f}(X)\Big) \; .
\end{equation}
This is actually given by the sum of the physical approximation error $\eta(Y^\star, Z)$ and another error term $Y - \hat{f}(X)$ which will be named the \emph{learning error}. Indeed it comes from the statistical approximation of $Y$ by $\hat{f}(X)$ and depends specifically on the training set $\mathcal{D}$, on the chosen model $f$ and the algorithm to compute the estimator.

\begin{lem} \label{lem:lem1}
    We have
    \begin{equation} \label{eq:err_bound}
        \E_{X,Y^\star}\Big[ \big| Y^\star - \hat{f}(X) \big| \Big] \leqslant r + \E_{X, Z} \Big[ \big| Y - \hat{f}(X) \big| \Big] \; .
    \end{equation}
\end{lem}
\begin{proof}
    By conditioning on $Z$, we have
    \begin{align}
        \E_{X,Y^\star}\Big[ \big| Y^\star - \hat{f}(X) \big| \Big]
            & = \E_{Z}\bigg[ \E_{X, Y^\star | Z} \Big[ \big| Y^\star - \hat{f}(X) \big| \, \Big| \, Z \Big] \bigg] \nonumber \\
            & \leqslant \E_{Z} \bigg[ \E_{X, Y^\star | Z} \Big[ \big| \eta(Y^\star,Z) \big| \, \Big| \, Z \Big] \bigg] + \E_{Z} \bigg[ \E_{X, Y^\star | Z} \Big[ \big| Y - \hat{f}(X) \big| \, \Big| \, Z \Big] \bigg] \label{eq:tri_ineq} \\
            & = \E_{Y^\star, Z}\Big[ \big| \eta(Y^\star,Z) \big| \Big] + \E_{X, Z} \Big[ \big| Y - \hat{f}(X) \big| \Big] \nonumber \\
            & \leqslant r + \E_{X, Z} \Big[ \big| Y - \hat{f}(X) \big| \Big]  \; ; \nonumber
    \end{align}
    note that we have used the triangle inequality applied to \eqref{eq:decomp_error} to obtain \eqref{eq:tri_ineq}.
\end{proof}

In the case where the order of magnitude of the learning error is smaller than the one of $r$, Lemma \ref{lem:lem1} shows in particular that trying to compute a more precise estimator will have little consequence on the above bound of the total error.

We end this section by comparing the total error with the mean value of $Y^\star$ in the following lemma. More precisely we upper bound the ratio between the means of the absolute value of the total error and of $Y^\star$ by an explicit and calculable quantity. This ratio, which can be reported as a percentage by multiplying it by 100, provides a relative measure of accuracy for the estimator $\hat{f}$. We also mention that it agrees with the Weighted Absolute Percentage Error (WAPE) in the classical case where $\hat{f}$ is an estimator for $Y^\star$.

\begin{lem} \label{lem:lem2}
    Suppose that $\E_Z\big[\varphi(Z) \big] > r$. Then $\E\big[Y^\star\big]$ is positive and we have
    \begin{equation} \label{eq:rel_err_bound}
         \frac{\E_{X,Y^\star}\Big[ \big| Y^\star - \hat{f}(X) \big| \Big]}{\E\big[Y^\star\big]} \leqslant \frac{r + \E_{X, Z} \Big[ \big| Y - \hat{f}(X) \big| \Big]}{\E_Z\big[ \varphi(Z) \big] - r} \; .
    \end{equation}
\end{lem}
\begin{proof}
    We have
    \begin{equation*}
        \E_{Y^\star, Z} \Big[ \varphi \big(Z \big) - Y^\star \Big] \leqslant \left| \E_{Y^\star, Z} \Big[ Y^\star - \varphi \big(Z \big) \Big] \right| \leqslant \E_{Y^\star, Z} \Big[ \big| Y^\star - \varphi \big(Z \big) \big| \Big] \; ,
    \end{equation*}
    where we have applied Jensen's inequality to the absolute value function to obtain the second inequality. Moreover the linearity of the expected value and the assumption
    \begin{equation*}
        \E_{Y^\star, Z} \Big[ \big| Y^\star - \varphi \big(Z \big) \big| \Big] \leqslant r
    \end{equation*}
    lead to
    \begin{equation} \label{eq:ineq_ref}
        \E_Z\big[ \varphi(Z) \big] - r \leqslant \E\big[ Y^\star \big] \; .
    \end{equation}
    Since $\E_Z\big[\varphi(Z) \big]$ is supposed to be larger than $r$, we deduce that $\E\big[Y^\star\big] > 0$. Then we can take the inverse of inequality \eqref{eq:ineq_ref} and combine the result with inequality \eqref{eq:err_bound} to obtain \eqref{eq:rel_err_bound}.
\end{proof}

In the following sections, we apply this abstract approach to model aerodynamic variables together with total error bounds. However it is noteworthy that this data-centric approach is sufficiently generic to be exploited in other disciplines.

\section{Application to aircraft performance} \label{sec:appl}

We now move to modelling the drag coefficient $C_D^\star$ and the lift coefficient $C_L^\star$ for a given narrow-body aircraft type for cruise conditions by exploiting recorded flight data\footnote{To be consistent with the notations introduced in Section \ref{sec:abs_setting}, we let $Y^\star$ denote an exact but unobserved variable and we define the observed variable $Y := Y^\star + \eta$, where $\eta$ denotes an error term.}. The predicted values of these coefficients are then expected to reflect real flights conditions. For instance, the coefficients $C_D^\star$ and $C_L^\star$ are used to establish the drag polar, which contains the aerodynamics of the aircraft \citep[Sec. 2.9]{A1999}. As it is classically assumed in the aeronautics literature \citep[see for instance][]{SHE2018}, our models for these coefficients will depend on the angle of attack $\alpha$ and on the Mach number $M$.

Nevertheless the coefficients $C_D^\star$ and $C_L^\star$ are neither observed nor measured by the sensors of the aircraft during the flight. We therefore leverage the approach developed in Section \ref{sec:abs_setting}. Following this approach, the main task is to determine approximated yet accurate formulas for the coefficients together with bounds for the physical approximation errors. With these approximations, we will be able to build data sets for approximated $C_D^\star$ and $C_L^\star$ which are expected to reflect on the actual aerodynamics of the aircraft. Models for $C_D^\star$ and $C_L^\star$ (depending on the angle of attack and Mach number) will then be trained on these data sets and their total errors will be bounded by using lemma \ref{lem:lem1}.

Prior to this, we emphasise that the method proposed in this paper is not limited to the present setting. It can be extended to other variables, aircraft types or phases, subject to available physical formulas and data.

For the sake of readability, Table \ref{table:units} provides the names, the symbols and the SI units of the main physical variables used in the rest of this paper. In addition the correspondence between the abstract variables and maps defined in Section \ref{sec:abs_setting} and the aeronautic ones is presented in Table \ref{table:correspondance}.

\begin{table}[ht]
    \centering
    \tabcolsep=5pt%
    \caption{Names, symbols and units of variables.}
    {\begin{tabular*}{.5\textwidth}{@{\extracolsep{\fill}}lll}
        \toprule
        Variable name & Symbol & Unit (SI) \\
        \midrule
        Angle of attack & $\alpha$ & rad \\
        Path angle & $\gamma$ & rad \\
        True airspeed & $V$ & m.s$^{-1}$ \\
        Mach number & $M$ & 1 \\
        Altitude & $h$ & m \\
        Mass & $m$ & kg \\
        Fuel flow & $\FF$ & kg.s$^{-1}$ \\
        Static air temperature & $\SAT$ & K\\
        Air density & $\rho$ & kg.m$^{-3}$ \\
        Thrust force & $T$ & N \\
        Drag force & $D$ & N \\
        Lift force & $L$ & N
    \end{tabular*}
    }
    \label{table:units}
\end{table}

\begin{table}[ht]
    \centering
    \tabcolsep=5pt%
    \caption{Correspondence between abstract variables defined in Section \ref{sec:abs_setting} and the physical variables.}
    {\begin{tabular*}{.45\textwidth}{@{\extracolsep{\fill}}cc}
        \toprule
        Abstract setting & Aeronautic setting \\
        \midrule
        $Y$ & $C_D$ or $C_L$ \\
        $X$ & $(\alpha, M)$ \\
        $Z$ & $(\rho, V, \alpha, \FF, \SAT, h, M, m, \gamma)$ \\
        $\varphi$ & See \eqref{eq:varphi_cd} and \eqref{eq:varphi_cl}
    \end{tabular*}
    }
    \label{table:correspondance}
\end{table}

Let us stress that the approximations we exploit here are actually derived from flight dynamics equations, whose accuracy depends on still existing physical models. In particular, we will use substantially the following approximated formula for the specific fuel consumption, noted here $C_{\SR}^\star$, from \citet[][Page 41]{Roux2005}:
\begin{align}
    C_{\SR}
        & := \Big( \big( a_1(h) \lambda + a_2(h) \big) M + \big(b_1(h) \lambda + b_2(h) \big) \Big) \sqrt{\frac{\SAT}{\SAT_0}} + \big( 7.4 \mathrm{e}{-13} (\varepsilon_c - 30) h + c \big) (\varepsilon_c - 30) \; , \label{eq:c_sp}
\end{align}
where
\begin{itemize}
    \item $\SAT_0$ is the temperature at sea level. Following the International Standard Atmosphere (ISA), it is set to 288.15 K;
	\item $\lambda$ is the bypass ratio which depends on the turbofan engines; here this value is fixed because we consider a single airliner type;
	\item $\varepsilon_c$ is the engine pressure ratio, which is also fixed here;
	\item $a_1, a_2, b_1, b_2$ are linear piecewise functions (depending on the altitude) and $c$ a constant which are given in \citet[Tab.~2.8]{Roux2005}.
\end{itemize}
As pointed out by \cite{Roux2002}, this model improves the classical one of \cite{T82} and its mean relative error and its standard deviation for cruise conditions are given in \citet[Page 66]{Roux2002}: they are equal respectively to 3.68\% and 4.48\%. Thus the coefficient $C_{\SR}^\star$ satisfies the following equation,
\begin{equation} \label{eq:eq_csp}
    C_{\SR}^\star = C_{\SR}(\SAT, h, M) + 
    \eta(C_{\SR}^\star, \SAT, h, M) 
\end{equation}
where
\begin{equation} \label{eq:error_mean}
    \E_{C_{\SR}^\star, \SAT, h, M} \left[ \frac{\big| \eta(C_{\SR}^\star, \SAT, h, M) \big|}{C_{\SR}^\star} \right] = 3.68 \times 10^{-2} \; ,
\end{equation}
over the cruise domain.

We establish now physical approximations for $C_D^\star$ and $C_L^\star$ in the case of a flight in a vertical plane and under the approximation that the Earth is locally flat. By applying Newton's second law to a body (modelling the aircraft) of mass $m$ moving in an air mass with no wind variations and by projecting the resulting equation onto the body frame, one obtains the following differential equations:
\begin{equation} \label{eq:diff_syst}
    \left \{ \begin{array}{l}
		m \, \dot{V} = T \cos \alpha - D - mg \sin \gamma \\[3mm]
		m \, V \, \dot{\gamma} = T \sin \alpha + L - m g \cos \gamma
    \end{array} \right. \; ,
\end{equation}
where $g$ is the value of gravitational acceleration on Earth (here rounded to 9.81 m.s$^{-2}$) and $\dot{x}$ denotes the time-derivative of any physical variable $x$. We refer for instance to \cite{Rommel2018} for a detailed derivation of the above relations. Moreover we have the following relations:
\begin{equation} \label{eq:rel_syst}
	\left\{ \begin{array}{l}
		\FF = C_{\SR}^\star \, T \\[2mm]
		\displaystyle D = \frac{1}{2} \, \rho \,  V^2 \, S \, C_D^\star \\[2mm]
		\displaystyle L = \frac{1}{2} \, \rho \,  V^2 \, S \, C_L^\star
	\end{array} \right. \; ,
\end{equation}
with $S$ denoting the wing-surface of the aircraft; note that this value is fixed in our setting. From the system \eqref{eq:rel_syst}, we clearly have
\begin{equation*}
    \left\{ \begin{array}{l}
		\displaystyle T = \frac{\FF}{C_{\SR}^\star} \\[2mm]
		\displaystyle C_D^\star = \frac{2}{\rho \,  V^2 \, S} \, D \\[2mm]
		\displaystyle C_L^\star = \frac{2}{\rho \,  V^2 \, S} \, L
	\end{array} \right. \; .
\end{equation*}
Combining the system \eqref{eq:diff_syst} with the preceding relations gives
\begin{equation} \label{eq:formulas_var}
    \left\{ \begin{array}{l}
		\displaystyle C_D^\star = \frac{2}{\rho \,  V^2 \, S} \, \left( \cos \alpha \, \frac{\FF}{C_{\SR}^\star} - m \, \dot{V} - mg \sin \gamma \right) \\[3mm]
		\displaystyle C_L^\star = \frac{2}{\rho \,  V^2 \, S} \, \left( -\sin \alpha \, \frac{\FF}{C_{\SR}^\star} + m \, V \, \dot{\gamma} + mg \cos \gamma \right)
	\end{array} \right. \; .
\end{equation}
Apart from the specific fuel consumption $C_{\SR}^\star$, all the variables appearing in the right-hand sides of the system \eqref{eq:formulas_var} are either recorded by the aircraft or easily calculable from other recorded variables via well-known physical relations.\\
By inserting equation \eqref{eq:eq_csp} into \eqref{eq:formulas_var}, we obtain the following formulas for $C_D^\star$ and $C_L^\star$:
\begin{equation*}
    \left\{ \begin{array}{l}
		\displaystyle T = \frac{\FF}{C_{\SR}} \, - \frac{\FF}{C_{\SR}} \, \frac{\eta_{C_{\SR}^\star}}{C_{\SR}^\star} \\[3mm]
		\displaystyle C_D^\star = \frac{2}{\rho \,  V^2 \, S} \, \left( \cos \alpha \, \frac{\FF}{C_{\SR}} - m \, \dot{V} - mg \sin \gamma - \cos \alpha \, \frac{\FF}{C_{\SR}} \,  \frac{\eta_{C_{\SR}^\star}}{C_{\SR}^\star} \right) \\[3mm]
		\displaystyle C_L^\star = \frac{2}{\rho \,  V^2 \, S} \, \left( -\sin \alpha \, \frac{\FF}{C_{\SR}} + m \, V \, \dot{\gamma} + mg \cos \gamma + \sin \alpha \, \frac{\FF}{C_{\SR}} \, \frac{\eta_{C_{\SR}^\star}}{C_{\SR}^\star} \right)
	\end{array} \right. \; , 
\end{equation*}
where $\eta_{C_{\SR}^\star} := \eta(C_{\SR}^\star, (\SAT, h, M))$ for the sake of simplicity. By defining
\begin{align}
    & \bullet \quad Z := (\rho, V, \alpha, \FF, \SAT, h, M, m, \gamma) \; ; \nonumber\\
    & \bullet \quad \varphi_{C_D^\star}\big(Z\big) := \frac{2}{\rho \,  V^2 \, S} \, \left( \cos \alpha \, \frac{\FF}{C_{\SR}(\SAT, h, M)} - m \, \dot{V} - mg \sin \gamma \right) \; ; \label{eq:varphi_cd} \\
    & \bullet \quad \eta_{C_D^\star}\big(C_D^\star, Z\big) := - \frac{2 \, \cos \alpha}{\rho \,  V^2 \, S} \, \frac{\FF}{C_{\SR}(\SAT, h, M)} \, \frac{\eta_{C_{\SR}^\star}}{C_{\SR}^\star} \; , \nonumber
\end{align}
we can write
\begin{equation*}
    C_D^\star = \varphi_{C_D^\star}\big(Z\big) + \eta_{C_D^\star}\big(C_D^\star, Z \big) \; ,
\end{equation*}
the variable $C_D := \varphi_{C_D^\star}\big(Z\big)$ being the desired approximation for $C_D^\star$.
Similarly we obtain
\begin{equation} \label{eq:cl}
    C_L^\star = \varphi_{C_L^\star}\big(Z\big) + \eta_{C_L^\star}\big(C_L^\star, Z\big) \; ,
\end{equation}
with
\begin{align}
    & \bullet \quad \varphi_{C_L^\star}\big(Z \big) := \frac{2}{\rho \,  V^2 \, S} \, \left( -\sin \alpha \, \frac{\FF}{C_{\SR}} + m \, V \, \dot{\gamma} + mg \cos \gamma \right) \; ; \label{eq:varphi_cl}\\
    & \bullet \quad \eta_{C_L^\star}\big(C_L^\star, Z \big) := \frac{2 \, \sin \alpha}{\rho \,  V^2 \, S} \, \frac{\FF}{C_{\SR}(\SAT, h, M)} \, \frac{\eta_{C_{\SR}^\star}}{C_{\SR}^\star} \; . \nonumber
\end{align}
Here the variable $C_L^\star$ is approximated by $C_L := \varphi_{C_L^\star}\big(Z\big)$.

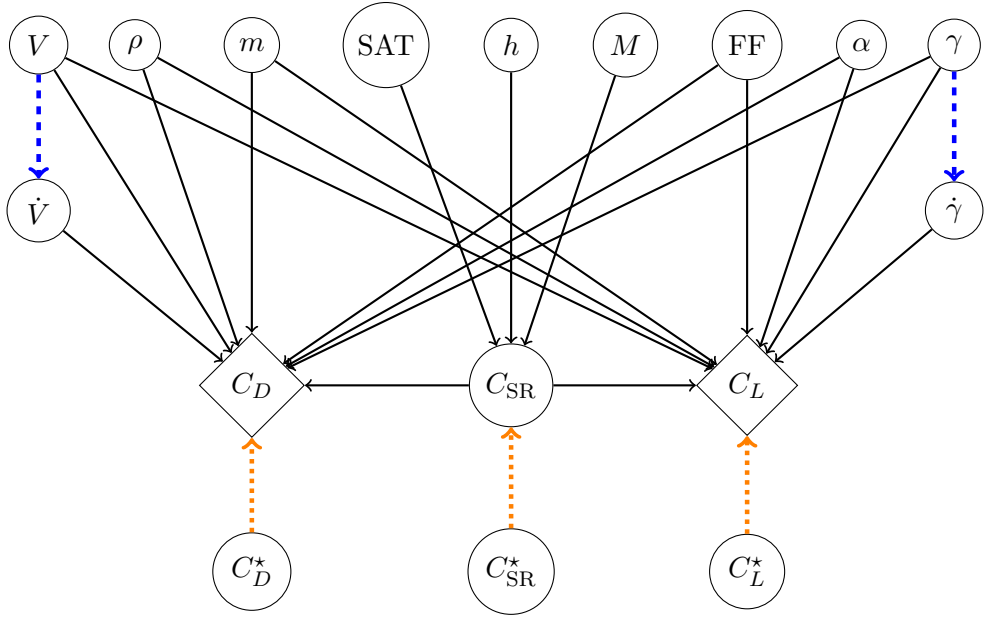
\begin{figure}[ht]
    \centering
    \begin{tikzpicture}
    \centering
    \matrix [column sep=5mm, row sep=12mm] {
        \node (v) [draw, shape=circle] {$V$}; &
        \node (rho) [draw, shape=circle] {$\rho$}; &
        \node (m) [draw, shape=circle] {$m$}; &
        \node (sat) [draw, shape=circle] {$\SAT$}; &
        \node (h) [draw, shape=circle] {$h$}; &
        \node (M) [draw, shape=circle] {$M$}; &
        \node (ff) [draw, shape=circle] {$\FF$}; &
        \node (aoa) [draw, shape=circle] {$\alpha$}; &
        \node (gam) [draw, shape=circle] {$\gamma$}; \\
        \node (vdot) [draw, shape=circle] {$\dot{V}$}; & & & & & & & &
        \node (gamdot) [draw, shape=circle] {$\dot{\gamma}$}; \\
        & & 
        \node (cdapp) [draw, shape=diamond] {$C_D$}; & &
        \node (cspapp) [draw, shape=circle] {$C_{\SR}$}; & &
        \node (clapp) [draw, shape=diamond] {$C_L$}; \\
        & &
        \node (cd) [draw, shape=circle] {$C_D^\star$}; & &
        \node (csp) [draw, shape=circle] {$C_{\SR}^\star$}; & &
        \node (cl) [draw, shape=circle] {$C_L^\star$}; \\
    };
    \draw[->, dashed, line width=.6mm, blue] (v) -- (vdot);
    \draw[->, dashed, line width=.6mm, blue] (gam) -- (gamdot);
    \draw[->, thick] (rho) -- (cdapp);
    \draw[->, thick] (v) -- (cdapp);
    \draw[->, thick] (aoa) -- (cdapp);
    \draw[->, thick] (ff) -- (cdapp);
    \draw[->, thick] (m) -- (cdapp);
    \draw[->, thick] (gam) -- (cdapp);
    \draw[->, thick] (vdot) -- (cdapp);
    \draw[->, thick] (cspapp) -- (cdapp);
    \draw[->, thick] (sat) -- (cspapp);
    \draw[->, thick] (h) -- (cspapp);
    \draw[->, thick] (M) -- (cspapp);
    \draw[->, thick] (rho) -- (clapp);
    \draw[->, thick] (v) -- (clapp);
    \draw[->, thick] (aoa) -- (clapp);
    \draw[->, thick] (ff) -- (clapp);
    \draw[->, thick] (m) -- (clapp);
    \draw[->, thick] (gam) -- (clapp);
    \draw[->, thick] (gamdot) -- (clapp);
    \draw[->, thick] (cspapp) -- (clapp);
    \draw[->, dotted, line width=.6mm, orange] (cd) -- (cdapp);
    \draw[->, dotted, line width=.6mm, orange] (csp) -- (cspapp);
    \draw[->, dotted, line width=.6mm, orange] (cl) -- (clapp);
    \end{tikzpicture}
    \caption{Relations between involved variables -- black arrows correspond to deterministic relations, differentiation with respect to time is represented by blue dashed arrows and the orange dotted ones refer to physical approximations; variables in diamond-shaped boxes are the targets we aim at modelling}
    \label{fig:diag}
\end{figure}

For the sake of readability, we sum up in Figure \ref{fig:diag} the relationships between the variables involved in the computations of $C_D$ and $C_L$. 

We now provide bounds for the means over the cruise phase of the absolute values of the physical approximation errors $\eta_{C_D^\star}$ and $\eta_{C_L^\star}$. Noting that these two variables are defined by a product in our setting, we can apply Hölder's inequality (with a choice of exponents $1$ and $+\infty$) to obtain
\begin{equation} \label{eq:error_bound_k}
    \left\{ \begin{array}{l}
       \displaystyle \E_{C_D^\star, Z} \Big[ \big| \eta_{C_D^\star}\big(C_D^\star, Z\big) \big| \Big] \leqslant K_{C_D^\star} \, \E_{C_{\SR}^\star, \SAT, h, M} \left[ \frac{\big| \eta(C_{\SR}^\star, \SAT, h, M) \big|}{C_{\SR}^\star} \right] \\[4mm]
       \displaystyle \E_{C_L^\star, Z} \Big[ \big| \eta_{C_D^\star}\big(C_D^\star, Z\big) \big| \Big] \leqslant K_{C_L^\star} \, \E_{C_{\SR}^\star, \SAT, h, M} \left[ \frac{\big| \eta(C_{\SR}^\star, \SAT, h, M) \big|}{C_{\SR}^\star} \right]
    \end{array} \right. \; ,
\end{equation}
where
\begin{align}
    &  K_{C_D^\star} := \sup_{(\rho, V, \alpha, \FF, \SAT, h, M, m)} \left| - \frac{2 \, \cos \alpha}{\rho \,  V^2 \, S} \, \frac{\FF}{C_{\SR}(\SAT, h, M)} \right| \; ; \label{eq:kcd}\\
    &  K_{C_L^\star} := \sup_{(\rho, V, \alpha, \FF, \SAT, h, M, m)} \left| \frac{2 \, \sin \alpha}{\rho \,  V^2 \, S} \, \frac{\FF}{C_{\SR}(\SAT, h, M)} \right| \; . \label{eq:kcl}
\end{align}
The supremum is over the cruise domain here and the mean absolute relative error for $C_{\SR}^\star$ is equal to $3.68\mathrm{e}{-2}$ according to \eqref{eq:error_mean}. We refer to Section \ref{sec:num_exp} for values of $K_{C_D^\star}$ and $K_{C_L^\star}$ computed from the available data.

\begin{remark} \label{rem:fss} \em
    Similarly to \cite{SHE2018}, it is possible to obtain simpler approximated formulas for $C_D^\star$ and $C_L^\star$ by assuming the following steady flight conditions with constant speed:
    \begin{enumerate}
        \item the altitude $h$ is constant and so the path angle $\gamma$ is equal to $0$;
        \item the angle of attack $\alpha$ is neglected: it is supposed to be equal to $0$;
        \item the true airspeed $V$ is constant and so its time-derivative $\dot{V}$ is equal to $0$.
    \end{enumerate}
    In this case, we have:
    \begin{equation*}
        \left\{ \begin{array}{l}
	    	\displaystyle T = \frac{\FF}{C_{\SR}} \, - \frac{\FF}{C_{\SR}} \, \frac{\eta_{C_{\SR}^\star}}{C_{\SR}^\star} \\[4mm]
		    \displaystyle C_D^\star = \frac{2}{\rho \,  V^2 \, S} \, \left( \frac{\FF}{C_{\SR}} - \frac{\FF}{C_{\SR}} \, \frac{\eta_{C_{\SR}^\star}}{C_{\SR}^\star} \right) \\[3mm]
		    \displaystyle C_L^\star = \frac{2 \, mg}{\rho \,  V^2 \, S}
	    \end{array} \right. \; .
    \end{equation*}
    In the present paper, we consider the complete formulas \eqref{eq:formulas_var} which are likely to best preserve accuracy of the approximations and to catch real flight conditions.
\end{remark}

\section{Experiments} \label{sec:num_exp}

In this section, we present numerical results based on real flight data for the method introduced in Section \ref{sec:abs_setting} and applied to the aeronautic setting described in Section \ref{sec:appl}. We first detail the data and the preprocessing steps we carried out, before reporting experiments design and results.

\subsection{Data description and preprocessing} \label{subsec:data_descript_preprocess}

We have access to 423 recorded short and medium-haul flights performed by the same narrow-body airliner, the data being recorded by the Quick Access Recorder (QAR). These flight data are  provided by a partner airline and can not be publicly released for commercial reasons. From this data set, we extract all the observations for the variables contained in the vector $Z$ defined in Section \ref{sec:appl}. The heading and the wind speed are also extracted to remove heading changes and high wind variations. All these variables are then smoothed by means of smoothing splines to remove the noise coming from measuring instruments and converted into the international system of units. The time-derivatives are computed on the basis of the smoothing splines. As explained in Section \ref{sec:appl}, we consider cruise phases in a vertical plane with no wind variations, so we require the following conditions to be satisfied:
\begin{itemize}
    \item we keep observations from the top of climb to the top of descent without those corresponding to climb steps; from a numerical point of view, we keep time-intervals such that the standard deviations of the altitude over these intervals is smaller than an arbitrary small threshold;
    \item the heading angle of the aircraft has to be constant; from a numerical point of view, we keep intervals such that the standard deviations of the heading over these intervals is smaller than an arbitrary small threshold; 
    \item the wind speed variations have to be equal to 0; from a numerical point of view, we keep intervals such that the means and the standard deviations of the time-derivative of the wind over these intervals are smaller than an arbitrary small threshold;
    \item the lengths of the resulting intervals have to be larger than 10 seconds;
\end{itemize}
Given the time-intervals during which the above conditions are satisfied, we sample every 10 seconds in each interval. This is motivated by the fact that the errors of the resulting models trained on the data set sampled every 10 seconds and on the data set without sampling are very close. Hence sampling allows to reduce the learning time without impacting strongly the accuracy. Afterwards the values for the approximated variables $C_D$ and $C_L$ are computed by means of the functions $\varphi_{C_D^\star}$ and $\varphi_{C_L^\star}$ defined in Section \ref{sec:appl}. Finally we have 164,054 observations which are randomly split into training, validation and test sets (70\% of the data set is used for the training, 20\% for the validation and 10\% for the test).

Table \ref{table:data} presents an example of a preprocessed data set (with simulated values to avoid divulgating the data set).

\begin{table}[ht]
    \tabcolsep=14.8pt%
    \caption{Example of a preprocessed data set.}
    {\begin{tabular*}{\textwidth}{@{\extracolsep{\fill}}|c|c|c|c|c|c|c|c|c|c|@{}}
        \hline
        Observation & $\rho$ & $V$ & $\alpha$ & $\FF$ & \ldots & $m$ & $\gamma$ \\
        \hline
        1 & 0.3224 & 234.5 & 0.0324 & 0.6716 & \ldots & 62,519 & 0.0139\\ 
        2 & 0.3704 & 236.8 & 0.0224 & 0.6503 & \ldots & 64,960 & 0.0198 \\
        3 & 0.3224 & 234.8 & 0.0305 & 0.6637 & \ldots & 66,974 & 0.0159 \\
        \vdots & \vdots & \vdots & \vdots & \vdots & & \vdots & \vdots \\
        164,054 & 0.3433 & 232.9 & 0.0332 & 0.6642 & \ldots & 66,673 & 0.0150 \\
        \hline
    \end{tabular*}
    }
    \label{table:data}
\end{table}

\subsection{Experiments design}

Here we aim at estimating the following models for the approximated drag $C_D$ and lift $C_L$ coefficients,
\begin{equation*}
    C_D = f_{C_D}(\alpha, M) \qquad \text{and} \qquad C_L = f_{C_L}(\alpha, M) \; ,
\end{equation*}
by exploiting the preprocessed data described above. To do so, we consider different classical models which are introduced in Table \ref{tab:models}. This table also gives the considered hyper-parameters and their range. The hyper-parameters are tuned by using 3-fold cross-validation, the loss function being the mean squared error. Furthermore we use an early stopping rule when fitting the gradient tree boosting model to limit the number of iterations, the validation set being used to stop iterating. The maximum number of iterations has been set to 5,000 in this case. In the end, we are interested in the three following classical error metrics: the root-square of the mean squared error (RMSE), the mean absolute error (MAE) and the mean absolute percent error (MAPE). We use the software package \texttt{LightGBM} \citep{Ke2017LightGBMAH} as an implementation of the gradient tree boosting algorithm and we compute the other models using \texttt{scikit-learn} Python library \citep{scikit-learn}.

\begin{table}[ht]
    \centering
    \tabcolsep=10pt%
    \caption{Hyper-parameters and their range for the considered models.}
    {\begin{tabular*}{.85\textwidth}{@{\extracolsep{\fill}}lllc@{}}
        \toprule
        Model & Hyper-parameters & Range \\
        \midrule
        {Constant} & none & $\emptyset$ \\
        {Linear} & none & $\emptyset$ \\
        Polynomial & degree & $\{2,3,4,5\}$ \\
        {SVM} & kernel & \{linear, polynomial, Gaussian, sigmoid\} \\
        {k-NN} & neighbours number, weights & $\{1, 21, 41, \dots, 701\} \times$\{uniform, distance\} \\
        {Decision tree} & trees depth & $\{1,2,\dots,10\}$ \\
        {Random forest} & trees depth, trees number & $\{1, 2, \dots, 6\} \times \{100, 200, \dots, 700\}$ \\
        {Gradient tree boosting} & trees depth & $\{1,2,\dots,6\}$ \\
    \end{tabular*}
    }
    \label{tab:models}
\end{table}

\subsection{Results}

We performed 100 times the learning process: at each time, the preprocessed data is randomly split into training, validation and test sets and the models are estimated and tested using these sets. The means and the standard deviations of the errors computed on the test sets are given in Tables \ref{table:errors_cd} and \ref{table:errors_cl}. We report numbers up to a precision of 3 decimal digits.

\begin{table}[ht]
    \tabcolsep=0pt%
    \caption{Means and standard deviations of error metrics for different $C_D$ models computed over 100 independent repetitions -- The smallest values are indicated by bolded numbers.}
    {\begin{tabular*}{\textwidth}{@{\extracolsep{\fill}}lccc@{}}
        \toprule
        $C_D$ model & RMSE & MAE & MAPE [\%] \\
        \midrule
        {Constant} & $( 8.778 \pm 0.285 ) \times 10^{-3}$ & $( 5.932 \pm 0.102 ) \times 10^{-3}$ & $53.58 \pm 56.58$ \\
        {Linear} & $ (1.992 \pm \mathbf{0.021} ) \times 10^{-3}$ & $( 1.424 \pm \mathbf{0.006} ) \times 10^{-3}$ & $4.53 \pm \mathbf{0.05}$ \\
        Polynomial & $(\mathbf{1.926} \pm 0.029) \times 10^{-3}$ & $( 1.361 \pm 0.008 ) \times 10^{-3}$ & $4.31 \pm 0.06$ \\
        {SVM} & $( 1.392 \pm 0.259 ) \times 10^{-2} $ & $(1.360 \pm 0.269 ) \times 10^{-2} $ & $43.25 \pm 8.86$ \\
        {k-NN} & $( 1.936 \pm 0.025 ) \times 10^{-3}$ & $( 1.362 \pm \mathbf{0.006}) \times 10^{-3}$ & $4.31 \pm \mathbf{0.05}$ \\
        {Decision tree} & $( 1.961 \pm 0.038 ) \times 10^{-3}$ & $( 1.364 \pm 0.009) \times 10^{-3}$ & $4.32 \pm \mathbf{0.05}$ \\
        {Random forest} & $(1.929 \pm 0.026) \times 10^{-3}$ & $( 1.357 \pm 0.007 ) \times 10^{-3}$ & $\mathbf{4.29} \pm \mathbf{0.05}$ \\
        {Gradient tree boosting} & $( 1.928 \pm 0.047 )\times 10^{-3}$ & $( \mathbf{1.356} \pm 0.012 ) \times 10^{-3}$ & $\mathbf{4.29} \pm 0.08$ \\
    \end{tabular*}
    }
    \label{table:errors_cd}
\end{table}

\begin{table}[ht]
    \tabcolsep=0pt%
    \caption{Means and standard deviations of error metrics for different $C_L$ models computed over 100 independent repetitions -- The smallest values are indicated by bolded numbers.}
    {\begin{tabular*}{\textwidth}{@{\extracolsep{\fill}}lccc@{}}
        \toprule
        $C_L$ model & RMSE & MAE & MAPE [\%] \\
        \midrule
        {Constant} & $( 7.358 \pm 0.116 ) \times 10^{-2}$ & $( 5.999 \pm 0.060 ) \times 10^{-2}$ & $14.24 \pm 0.14$ \\
        {Linear} & $ (1.436 \pm 0.007 ) \times 10^{-2}$ & $( 1.102 \pm 0.004 ) \times 10^{-2}$ & $2.17 \pm \mathbf{0.01}$ \\
        Polynomial & $(1.205 \pm \mathbf{0.005}) \times 10^{-2}$ & $( 9.203 \pm \mathbf{0.036} ) \times 10^{-3}$ & $1.78 \pm \mathbf{0.01}$ \\
        {SVM} & $( 6.272 \pm 0.029 ) \times 10^{-2} $ & $(5.982 \pm 0.030 ) \times 10^{-2} $ & $11.16 \pm 0.05$ \\
        {k-NN} & $( \mathbf{1.181} \pm 0.006 ) \times 10^{-2}$ & $( 8.976 \pm 0.039) \times 10^{-3}$ & $1.73 \pm \mathbf{0.01}$ \\
        {Decision tree} & $( 1.185 \pm 0.008 ) \times 10^{-2}$ & $( \mathbf{8.886} \pm 0.045) \times 10^{-3}$ & $\mathbf{1.71} \pm \mathbf{0.01}$ \\
        {Random forest} & $(1.209 \pm 0.006) \times 10^{-2}$ & $( 9.211 \pm 0.041) \times 10^{-3}$ & $1.78 \pm \mathbf{0.01}$ \\
        {Gradient tree boosting} & $( 1.192 \pm 0.011 )\times 10^{-2}$ & $( 9.025 \pm 0.072 ) \times 10^{-3}$ & $1.75 \pm 0.02$ \\
    \end{tabular*}
    }
    \label{table:errors_cl}
\end{table}

\begin{figure}[ht]
    \centering
    \includegraphics[width=\textwidth]{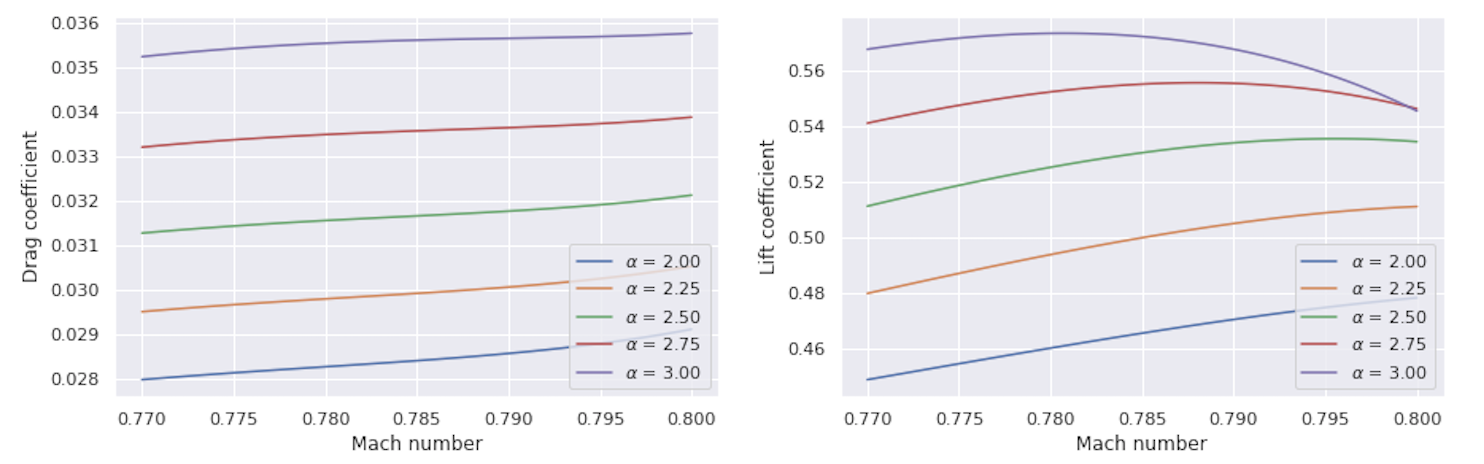}
    \caption{Predictions of $C_D$ and $C_L$ from polynomial models}
    \label{fig:polar_poly}
\end{figure}

\begin{figure}[ht]
    \centering
    \includegraphics[width=\textwidth]{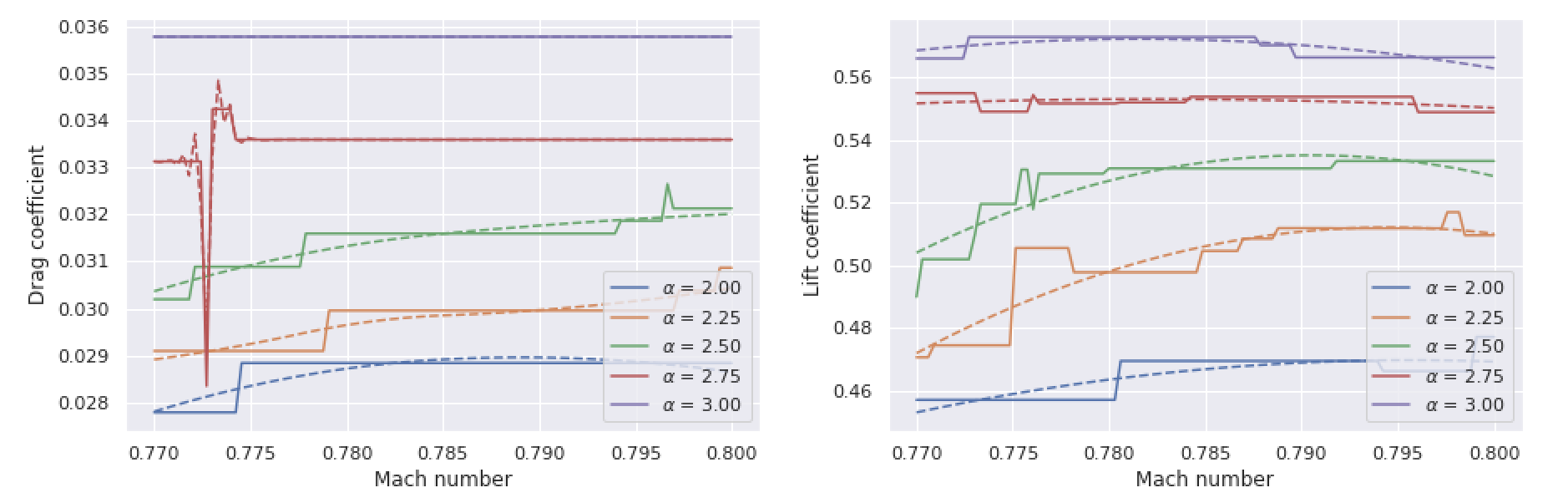}
    \caption{Predictions of $C_D$ and $C_L$ from decision trees models --  Solid lines are the raw prediction curves and dotted lines are smoothed versions (using splines)}
    \label{fig:polar_tree}
\end{figure}

\begin{figure}[ht]
    \centering
    \includegraphics[width=\textwidth]{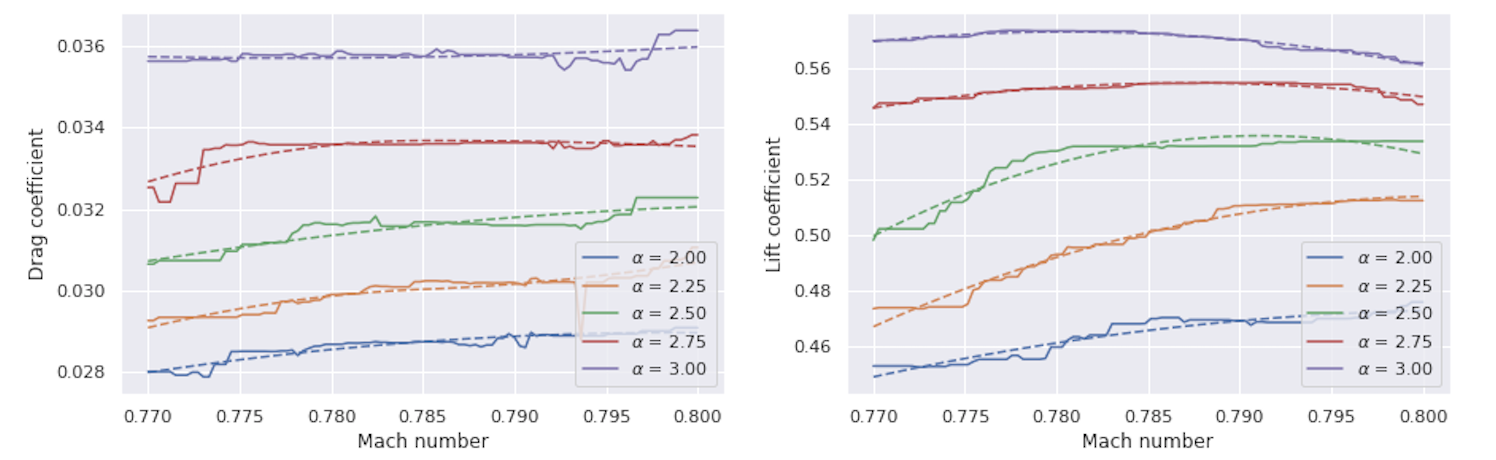}
    \caption{Predictions of $C_D$ and $C_L$ from decision gradient tree boosting models -- Solid lines are the raw prediction curves and dotted lines are smoothed versions (using splines)}
    \label{fig:polar_lgbm}
\end{figure}

Figures \ref{fig:polar_poly}, \ref{fig:polar_tree} and \ref{fig:polar_lgbm} allow to visualise the tendencies of estimators $\hat{f}_{C_D}$ and $\hat{f}_{C_L}$ with respect to the Mach number for different fixed values of the angle of attack. Figures \ref{fig:polar_poly}, \ref{fig:polar_tree} and \ref{fig:polar_lgbm} show respectively a polynomial, a decision tree and a gradient tree boosting models.

First of all we observe that the decision tree and gradient tree boosting models lead to raw predicted curves which may be hard to interpret from an aeronautic point of view. This is especially true for the decision tree model even though its learning error is similar to those of the two other models. Since we aim here at checking whether some expected aeronautic tendencies are caught by our approach, we smooth the predicted curves by means of smoothing splines to interpret the results in an easier way. These smoothed curves are given by the dotted curves in Figures \ref{fig:polar_tree} and \ref{fig:polar_lgbm}. Note that the use of smoothing splines is to highlight the underlying tendency, but is certainly not used as a predictive method because of well-known artefacts due to the smoothing technique.

Now we mention that 90\% of Mach number data are between 0.77 and 0.80 and 90\% of angle of attack data are between 1.9$^{\circ}$ and 2.9$^{\circ}$. Then we observe that both predicted $C_D$ and $C_L$ globally increase when the Mach number or the angle of attack increases. This global tendency is actually expected in this small range of values according to \citet[Part 1, Chap. 2]{A1999}: the larger the angle of attack or the Mach number, the larger the drag and lift coefficients.

Nevertheless, this natural tendency for the lift coefficient is not verified by the estimators when $\alpha$ is too large, namely $\alpha=2.75^{\circ}$ or $\alpha=3^{\circ}$. This unexpected behaviour can be explained by the approximated nature of the variable $C_L$. Indeed it may behave in a way that is different from $C_L^\star$ in certain regions of the cruise domain. In this case, any estimator for $C_L$ is likely to inherit this unexpected behaviour and we believe refined aeronautics-supported approximations would bring a solution.

We now focus on the physical approximation errors for the drag and lift coefficients, that is to say $\eta_{C_D^\star}$ and $\eta_{C_L^\star}$. According to \eqref{eq:error_bound_k}, the mean of the absolute value of these errors is bounded by the product between the constants $K_{C_D^\star}$ or $K_{C_L^\star}$ with the mean absolute relative error of $C_{\SR}^\star$, the latter being equal to $3.68\times 10^{-2}$. We estimate the value of these constants by using our recorded observations: the maximal values of $K_{C_D^\star}$ or $K_{C_L^\star}$ (defined in Eqs. \eqref{eq:kcd} and \eqref{eq:kcl}) are respectively equal to $4.38\times 10^{-2}$ and $2.94\times 10^{-3}$. Using these two values as estimators for $K_{C_D^\star}$ and $K_{C_L^\star}$ gives the following bounds for the physical approximation errors:
\begin{equation*}
    \left\{ \begin{array}{l}
       \displaystyle \E_{C_D^\star, Z} \Big[ \big| \eta_{C_D^\star}\big(C_D^\star, Z\big) \big| \Big] \leqslant 1.61\times 10^{-3} \\[3mm]
       \displaystyle \E_{C_L^\star, Z} \Big[ \big| \eta_{C_L^\star}\big(C_L^\star, Z\big) \big| \Big] \leqslant 1.08\times 10^{-4}
    \end{array} \right. \; .
\end{equation*}

According to the generic inequality given in Lemma \ref{lem:lem1}, a bound for the mean of the absolute total error (defined in Section \ref{sec:abs_setting}) of a given variable can be obtained by adding up the bounds for the physical approximation and learning errors. The latter is here approximated by the MAE of the estimated model. To provide an example of numerical bounds for the total errors of the drag and lift coefficients, we choose estimators $\hat{f}_{C_D}$ and $\hat{f}_{C_L}$ whose MAE values are equal to the MAE means given in Tables \ref{table:errors_cd} and \ref{table:errors_cl}. Note that this choice is motivated by the fact that the standard deviations of the MAE for the different models are much smaller than the means. Following this choice, examples of bounds for the total errors are then given in Table \ref{table:total_errors}.

The empirical means of $C_D = \varphi_{C_D^\star}(Z)$ and $C_L = \varphi_{C_L^\star}(Z)$ over our data are respectively equal to
\begin{equation*}
    \widehat{\E_Z}\big[ \varphi_{C_D^\star}(Z) \big] = 3.23\times 10^{-2}\ , \qquad \widehat{\E_Z}\big[ \varphi_{C_L^\star}(Z) \big] = 5.32\times 10^{-1} \; ,
\end{equation*}
showing in particular that $\widehat{\E_Z}\big[ \varphi_{C_D^\star}(Z) \big] > 1.61\times 10^{-3}$ and $\widehat{\E_Z}\big[ \varphi_{C_L^\star}(Z) \big] > 1.08\times 10^{-4}$ (we used the slight notation abuse $\widehat{\E}$ to denote the empirical mean). The hypotheses of Lemma \ref{lem:lem2} are then satisfied and we are in position to upper bound the ratios between the mean of the absolute value of the total errors and the mean values of $C_D^\star$ and $C_L^\star$. Numerical values expressed as a percentage for these bounds are given in Table \ref{table:total_errors}.

\begin{table}[ht]
    \centering
    \tabcolsep=0pt%
    \caption{Bounds for the mean absolute and mean relative total errors of the drag and lift coefficients using estimators $\hat{f}_{C_D}$ and $\hat{f}_{C_L}$ whose MAE values are equal to the MAE means given in Tables \ref{table:errors_cd} and \ref{table:errors_cl} -- \emph{Absolute} and \emph{Relative} refer respectively to the bounds given in Lemmas \ref{lem:lem1} and \ref{lem:lem2}.}
    {\begin{tabular*}{.85\textwidth}{@{\extracolsep{\fill}}lcccc@{}}
        \toprule
        & \multicolumn{2}{@{}c@{}}{Drag coefficient} & \multicolumn{2}{@{}c@{}}{Lift coefficient} \\
        \cmidrule{2-3}\cmidrule{4-5}
        Models & Absolute & Relative [$\%$] & Absolute & Relative [$\%$] \\
        \midrule
        Constant & $7.542\times 10^{-3}$ & 24.57 & $6.010\times 10^{-2}$ & 11.30 \\
        Linear & $3.034\times 10^{-3}$ & 9.89 & $1.113\times 10^{-2}$ & 2.09 \\
        Polynomial & $2.971\times 10^{-3}$ & 9.68 & $9.311\times 10^{-3}$ & 1.75 \\
        SVM & $1.521\times 10^{-2}$ & 49.56 & $5.993\times 10^{-2}$ & 11.27 \\
        k-NN & $2.972\times 10^{-3}$ & 9.68 & $9.084\times 10^{-3}$ & 1.71 \\
        Decision tree & $2.974\times 10^{-3}$ & 9.69 & $\mathbf{8.994\times 10^{-3}}$ & \textbf{1.69} \\
        {Random forest} & $2.967\times 10^{-3}$ & \textbf{9.67} & $9.319\times 10^{-3}$ & 1.75 \\
        Gradient tree boosting & $\mathbf{2.966\times 10^{-3}}$ & \textbf{9.67} & $9.133\times 10^{-3}$ & 1.72 \\
    \end{tabular*}
    }
    \label{table:total_errors}
\end{table}

These results capture how classical, off-the-shelf regression methods perform and somewhat surprisingly, most methods compete on similar grounds (to the notable exception of the SVM). This suggests that most of these methods achieve a good enough complexity to capture the underlying phenomenon.

To finish, we mention that our approach is sufficiently generic to be applied to other settings, such as other flight phases. For the sake of illustration, we consider the drag coefficient during the climb phase. In this case, we exploit the climb data from our 423 available recorded flights, that is to say data for which the altitude is between 3,000 ft and the top of climb, and we apply the same preprocessing and learning steps as those described in this section. The results for the means and standard deviations of the learning errors computed over a test set are given in Table \ref{table:errors_cd_climb}. We remark that these errors are much larger than those for the cruise phase. This accuracy loss can be explained by the fact that each variable during the climb phase has a larger range. For instance, the Mach number varies from 0.3 at the beginning of the climb to 0.81 at the top of climb while it varies only from 0.76 and 0.81 during the cruise. In addition, we are not in position to compute a numerical value for a bound of the physical approximation errors means for $C_D^\star$. This is due to the fact that \cite{Roux2002} does not estimate the physical approximation error mean coming from its model for the variable $C_{\SR}^\star$ for the climb phase. Once such a quantity is available, numerical bounds for the total errors for $C_D^\star$ can be derived in this case.

\begin{table}[ht]
    \tabcolsep=0pt%
    \caption{Means and standard deviations of error metrics for different $C_D$ models for climb phase computed over 100 independent repetitions -- The smallest values are indicated by bolded numbers.}
    {\begin{tabular*}{\textwidth}{@{\extracolsep{\fill}}lccc@{}}
        \toprule
        $C_D$ model & RMSE & MAE & MAPE [\%] \\
        \midrule
        {Constant} & $( 8.912 \pm 0.373 ) \times 10^{-3}$ & $( 5.919 \pm 0.102 ) \times 10^{-3}$ & $48.21 \pm 40.71$ \\
        {Linear} & $ (6.061 \pm 0.208 ) \times 10^{-3}$ & $( 3.618 \pm 0.041 ) \times 10^{-3}$ & $29.08 \pm \mathbf{14.16}$ \\
        Polynomial & $(5.611 \pm 0.215) \times 10^{-3}$ & $( \mathbf{3.399} \pm 0.037 ) \times 10^{-3}$ & $27.94 \pm 14.33$ \\
        {SVM} & $( 2.274 \pm 0.708 ) \times 10^{-2} $ & $(1.937 \pm 0.659 ) \times 10^{-2} $ & $82.99 \pm 24.74$ \\
        {k-NN} & $( 5.639 \pm 0.215 ) \times 10^{-3}$ & $( 3.420 \pm \mathbf{0.035}) \times 10^{-3}$ & $26.90 \pm 14.93$ \\
        {Decision tree} & $( 5.840 \pm 0.315 ) \times 10^{-3}$ & $( 3.494 \pm 0.050) \times 10^{-3}$ & $\mathbf{26.74} \pm 14.26$ \\
        {Random forest} & $(\mathbf{5.598} \pm \mathbf{0.189}) \times 10^{-3}$ & $( 3.428 \pm \mathbf{0.035} ) \times 10^{-3}$ & $27.02 \pm 16.70$ \\
        {Gradient tree boosting} & $( 5.678 \pm 0.425 )\times 10^{-3}$ & $( 3.405 \pm 0.073 ) \times 10^{-3}$ & $33.27 \pm 32.07$ \\
    \end{tabular*}
    }
    \label{table:errors_cd_climb}
\end{table}

\section{Conclusion}\label{sec:conclusion}

Our contributions are twofold: (i) we have proposed individual models trained on in-air data to improve the current aeronautic performance of individual aircrafts, rather than industry-wide calibrated parameters. This allows in particular for the search of more efficient (\emph{e.g.}, flight duration, speed, fuel consumption, etc.) trajectories for aircrafts (ii) we have designed a generic framework combining off-the-shelf machine learning with domain-specific approximations, which can be used in any data-intensive engineering discipline. We certainly hope that this approach can be replicated in other fields of study. We also intend to use this approach as a building block to optimising end-to-end pipelines, \emph{e.g.} for in-air real-time fuel optimisation.

\paragraph{Acknowledgments}
The authors are grateful to Baptiste Gregorutti and Pierre Jouniaux for fruitful discussions about the conceptualisation of this research and the aeronautic validation of the results. The authors are grateful to Arthur Talpaert for his substantial contribution to the code used to implement our methods.

\paragraph{Funding statement}
This project has received funding from the Clean Sky 2 Joint Undertaking under the European Union’s Horizon 2020 research and innovation programme under grant agreement No 815914 (PERF-AI). The funder had no role in study design, data collection and analysis, decision to publish, or preparation of the manuscript.

\bibliography{biblio}

\begin{thebibliography}{17}
\providecommand{\natexlab}[1]{#1}
\providecommand{\url}[1]{\texttt{#1}}
\expandafter\ifx\csname urlstyle\endcsname\relax
  \providecommand{\doi}[1]{doi: #1}\else
  \providecommand{\doi}{doi: \begingroup \urlstyle{rm}\Url}\fi

\bibitem[Airbus(2001)]{airbus2001}
Airbus.
\newblock Getting hands on experience with aerodynamic deterioration.
\newblock Technical report, Airbus, 2001.
\newblock
  http://www.smartcockpit.com/docs/Getting\_Hands-On\_Experience\_With\_Aerodynamic\_Deteriorations.pdf.

\bibitem[Airbus(2002)]{airbus2002}
Airbus.
\newblock Getting to grips with aircraft performance monitoring.
\newblock Technical report, Airbus, 2002.
\newblock
  http://www.smartcockpit.com/docs/Getting\_to\_Grips\_With\_Aircraft\_Performance.pdf.

\bibitem[Anderson(1999)]{A1999}
J.D. Anderson.
\newblock \emph{Aircraft performance and design}.
\newblock McGraw-Hill international editions: Aerospace science/technology
  series. WCB/McGraw-Hill, 1999.

\bibitem[Buonaccorsi(1996)]{Buonaccorsi1996}
John~P. Buonaccorsi.
\newblock Measurement error in the response in the general linear model.
\newblock \emph{Journal of the American Statistical Association}, 91\penalty0
  (434):\penalty0 633--642, 1996.
\newblock ISSN 01621459.
\newblock URL \url{http://www.jstor.org/stable/2291659}.

\bibitem[Carroll and Ruppert(1988)]{carroll1988transformation}
Raymond~J Carroll and David Ruppert.
\newblock \emph{Transformation and weighting in regression}, volume~30.
\newblock CRC Press, 1988.

\bibitem[Fuller(2009)]{fuller2009}
Wayne~A Fuller.
\newblock \emph{Measurement error models}, volume 305.
\newblock John Wiley \& Sons, 2009.

\bibitem[Ke et~al.(2017)Ke, Meng, Finley, Wang, Chen, Ma, Ye, and
  Liu]{Ke2017LightGBMAH}
Guolin Ke, Qi~Meng, Thomas Finley, Taifeng Wang, Wei Chen, Weidong Ma, Qiwei
  Ye, and Tie-Yan Liu.
\newblock Lightgbm: A highly efficient gradient boosting decision tree.
\newblock In I.~Guyon, U.~V. Luxburg, S.~Bengio, H.~Wallach, R.~Fergus,
  S.~Vishwanathan, and R.~Garnett, editors, \emph{Advances in Neural
  Information Processing Systems 30}, pages 3146--3154. Curran Associates,
  Inc., 2017.

\bibitem[Krajček et~al.(2015)Krajček, Nikolić, and Domitrović]{krajcek2015}
K~Krajček, D~Nikolić, and A~Domitrović.
\newblock Aircraft performance monitoring from flight data.
\newblock \emph{Tehnički vjesnik}, 22\penalty0 (5):\penalty0 1337--1344, 2015.
\newblock \doi{https://doi.org/10.17559/TV-20131220145918}.

\bibitem[Loftin(1985)]{L1985}
L.~K. Loftin.
\newblock \emph{Quest for performance: The evolution of modern aircraft}.
\newblock NASA Scientific and Technical Information Branch, 1985.

\bibitem[McCormick(1995)]{MC1995}
B.~W. McCormick.
\newblock \emph{Aerodynamics, Aeronautics, and Flights Mechanics, Second
  Edition}.
\newblock Wiley, 1995.

\bibitem[Pedregosa et~al.(2011)Pedregosa, Varoquaux, Gramfort, Michel, Thirion,
  Grisel, Blondel, Prettenhofer, Weiss, Dubourg, Vanderplas, Passos,
  Cournapeau, Brucher, Perrot, and Duchesnay]{scikit-learn}
F.~Pedregosa, G.~Varoquaux, A.~Gramfort, V.~Michel, B.~Thirion, O.~Grisel,
  M.~Blondel, P.~Prettenhofer, R.~Weiss, V.~Dubourg, J.~Vanderplas, A.~Passos,
  D.~Cournapeau, M.~Brucher, M.~Perrot, and E.~Duchesnay.
\newblock Scikit-learn: Machine learning in {P}ython.
\newblock \emph{Journal of Machine Learning Research}, 12:\penalty0 2825--2830,
  2011.

\bibitem[Rommel(2018)]{Rommel2018}
C.~Rommel.
\newblock \emph{Exploration de données pour l'optimisation de trajectoires
  aériennes}.
\newblock PhD thesis, \'Ecole Polytechnique, 2018.

\bibitem[Roux(2002)]{Roux2002}
\'E Roux.
\newblock \emph{Modèles Moteurs... Réacteurs double flux civils et réacteurs
  militaires à faible taux de dilution avec Post-Combustion}.
\newblock INSA-SupAéro-ON\'ERA, 2002.

\bibitem[Roux(2005)]{Roux2005}
\'E. Roux.
\newblock \emph{Pour une approche analytique de la Dynamique du Vol}.
\newblock PhD thesis, SupAéro, 2005.

\bibitem[Schennach(2016)]{SCH2016}
Susanne~M. Schennach.
\newblock Recent advances in the measurement error literature.
\newblock \emph{Annual Review of Economics}, 8\penalty0 (1):\penalty0 341--377,
  2016.
\newblock \doi{10.1146/annurev-economics-080315-015058}.
\newblock URL \url{https://doi.org/10.1146/annurev-economics-080315-015058}.

\bibitem[Sun et~al.(2018)Sun, Hoekstra, and Ellerbroek]{SHE2018}
J.~Sun, J.~M. Hoekstra, and J.~Ellerbroek.
\newblock Aircraft drag polar estimation based on a stochastic hierarchical
  model.
\newblock In \emph{Eighth SESAR Innovation Days}, 2018.

\bibitem[Torenbeek(1982)]{T82}
E.~Torenbeek.
\newblock \emph{Synthesis of subsonic airplane design}.
\newblock Delft University Press, 1982.

\end{thebibliography}

\end{document}